\documentclass[USenglish]{article}	
\usepackage[utf8]{inputenc}				%(only for the pdftex engine)
\usepackage[big,online]{dgruyter}	%values: small,big | online,print,work
\usepackage{lmodern} 
\usepackage{microtype}
\usepackage[justification=centering]{caption}
\usepackage[numbers,square,sort&compress]{natbib}
\usepackage{color, colortbl,booktabs}
\definecolor{Gray}{gray}{0.9}

\theoremstyle{dgthm}

\theoremstyle{dgdef}

\begin{document}

%%%--------------------------------------------%%%
	\articletype{Research Article}
	\received{Month	DD, YYYY}
	\revised{Month	DD, YYYY}
  \accepted{Month	DD, YYYY}
  \journalname{J.~Quant.~Anal.~Sports}
  \journalyear{YYYY}
  \journalvolume{XX}
  \journalissue{X}
  \startpage{1}
  \aop
  \DOI{10.1515/sample-YYYY-XXXX}
%%%--------------------------------------------%%%

\title{An analysis of the NCAA college football playoff team selections using an Elo ratings model}
\runningtitle{An analysis of the CFP selections using Elo ratings}
%\subtitle{Insert subtitle if needed}

\author*[1]{Benjamin Lucas}
%\ use * to mark the author as the corresponding author
\runningauthor{B.~Lucas}
\affil[1]{\protect\raggedright 
Northern Arizona University, Department of Mathematics and Statistics,  Flagstaff, AZ, United States e-mail: ben.lucas@nau.edu}
% \affil[2]{\protect\raggedright 
% Institution, Department, City, Country of second author and third author, e-mail: author\_two@xx.yz, author\_three@xx.yz}
	
%\communicated{...}
%\dedication{...}
	
\abstract{
In December 2023 the Florida State Seminoles became the first Power 5 school to have an undefeated season and miss selection for the College Football Playoff. In order to assess this decision, we employed an Elo ratings model to rank the teams and found that the selection committee's decision was justified and that Florida State were not one of the four best teams in college football in that season (ranking only $11^{th}$!). We extended this analysis to all other years of the CFP and found that the top four teams by Elo ratings differ greatly from the four teams selected in almost every year of the CFP's existence. Furthermore, we found that there have been more egregious non-selections including when Alabama was ranked first by Elo ratings in 2022 and were not selected. The analysis suggests that the current criteria are too subjective and a ratings model should be implemented to provide transparency for the sport, its teams, and its fans.
}


\maketitle

\section{Introduction} 
On December 2nd 2023, the Florida State Seminoles football team won the Atlantic Coast Conference championship game against the University of Louisville Cardinals, finishing off an undefeated season of 13 wins and 0 losses.
Despite this, the following day they were not selected by the College Football Playoff (CFP) selection committee as one of four teams to compete for the college football national championship in January.
This decision rendered them the first team from a Power 5 conference\footnote{The Power 5 are: Atlantic Coast Conference (ACC), Big Ten Conference, Big 12 Conference, Pac-12 Conference, and Southeastern Conference (SEC).}to record an undefeated season and miss selection for the CFP. The committee's decision is an openly subjectively exercise and this year (like many years) was met with great criticism from national sports media~\cite{espnFloridaState, cnnFloridaState, cbssportsFloridaState} (whose opinions are often similarly subjective), but one cannot help but wonder whether the decision could be adequately justified.

The CFP committee chair Boo Corrigan explained to a national media organization that Florida State was left out based on the ``unavailability of key players'' \cite{espnFloridaState}, which is one of many criteria that the committee uses for selection. This comment was made in reference to the injury to star quarterback Jordan Travis, who suffered a season-ending leg injury in the game against North Alabama on November 18, 2023. It could be argued however, that Florida State were a great team regardless the Travis injury, as evidenced by them winning their final two games of the regular season, including the ACC championship game against the highly-ranked University of Louisville.

In this paper, we seek to quantitatively answer the question of whether the selection committee was correct in excluding Florida State from the CFP using Elo rankings--a method for calculating the relative skill of competitors in head-to-head zero-sum games (such as American football).

\section{Background}
Since 2014, at the conclusion of the conference championship games, four college football teams from Division 1 of the National Collegiate Athletic Association (NCAA) have been invited to compete in the CFP, a knockout tournament used to award a national champion. The tournament consists of three games: two semi-finals--where the selected teams ranked 1 and 4 play against each other, and so do the teams ranked 2 and 3--followed by a championship game between the winners of the two semi-finals\footnote{We note here that the CFP occurs in the year following the season it is concluding. To avoid confusing we refer the CFP by the year that the regular season occurred--for example the 2021 CFP, concluded the 2021 season, but actually occurred in January 2022.}.

\subsection{CFP selection committee and criteria}
The CFP selection committee consists of 13 members serving three-year terms. The committee members include a current athletic director from each of Power 5 conferences; former athletic directors, coaches, players and administrators; and a former media member. The committee is also selected with the intention of geographical balance~\cite{committee}.

The teams invited by the selection committee to participate in the CFP are selected using the following criteria~\cite{criteria}:
\begin{enumerate}
    \item Conference championships won;
    \item Strength of Schedule;
    \item Head-to-head games between teams vying for selection;
    \item Comparative outcomes of common opponents; and,
    \item Other relevant factors such as unavailability of key players and coaches that may have affected a team’s performance during the season or likely will affect its postseason performance. 
\end{enumerate}

The final point here was the criterion used to exclude Florida State, as quarterback Jordan Travis was unavailable for the 2023 CFP due to an injury suffered in Week 13.

\subsection{CFP selection statistics}

Table~\ref{team_selections} shows the teams selected for the CFP since 2014. The teams who have appeared multiple times would not be a surprise to any college football fan as these are historically dominant programs, with Alabama appearing 8 times, which is every year bar one. Alabama, Clemson, and Georgia account for seven of the nine championships awarded.

\begin{table}[htp!]
\centering \caption{The number of CFP selections and national championships by team.}
\begin{tabular}{c|c|c}
Team & Selections & Championships Won \\
\hline
Alabama & 8 & 3 \\
Clemson & 6 & 2 \\
Ohio State & 5 & 1 \\
Oklahoma & 4 & 0 \\
Georgia & 3 & 2 \\
Michigan & 3 & 0 \\
Notre Dame & 2 & 0 \\
Washington & 2 & 0 \\
LSU & 1 & 1 \\
Oregon & 1 & 0 \\
TCU & 1 & 0 \\
Florida State & 1 & 0 \\
Michigan State & 1 & 0 \\
Cincinnati & 1 & 0 \\
Texas & 1 & 0 \\
\end{tabular}
\label{team_selections}
\end{table}

Table~\ref{conference_selections} shows the number of times teams from each conference have been selected, with the Power 5 conferences being clearly dominant.

\begin{table}[htp!]
\centering \caption{The number of CFP selections by conference.}
\begin{tabular}{c|c|c}
Conference & Selections & No. of teams \\
SEC & 12 & 3 \\
Big Ten & 9 & 3 \\
ACC & 8 & 3 \\
Big 12 & 6 & 2 \\
Pac-12 & 3 & 2 \\
Independent & 1 & 1 \\
American & 1 & 1 \\
\end{tabular}
\label{conference_selections}
\end{table}

\section{Elo ratings}
Elo Ratings were created by physicist--and chess master--Arpad Elo in the late 1950s, and were originally developed to rank chess players (and are still used for this purpose). Elo Ratings are used to define a relative skill level of a player in head-to-head zero-sum games. While their origin is in chess, Elo ratings are now applied across many sports and games including soccer \cite{arntzen2021predicting, hvattum2010using, gasquez2016determinants}, snooker \cite{collingwood2022evaluating}, hockey \cite{tenkanen2019rating}, Australian football \cite{ryall2010optimized}, and in professional American football \cite{lee2018testing, ziemba20152014, dabadghao2022predictive}.

Over many decades Elo ratings have proven to be a fair and comprehensive ranking method~\cite{tsang2016fabric, gasquez2016determinants}. Additionally, it has been shown to have great predictive power, consistently predicting the winners and close margins for a variety of sports~\cite{angelini2022weighted, vaughan2021well, tenkanen2019rating}.

Elo ratings have proven so reliable that they have even been adapted for applications outside of sports and games. For instance, \citet{pelanek2016applications} use the system in the education setting where the head-to-head competition is between the student's answer and the item.

\subsection{How are they calculated?}
Elo ratings work by comparing the opposing teams' current ratings and setting up an expected outcome of the match. Once the match is completed, the rating for each side is updated by comparing the actual outcome to the expected outcome based on Elo. As the ratings are zero-sum, the magnitude of the increase in one team's rating is equal to the magnitude of the decrease in the other team's rating.

To create Elo ratings, each team is given a starting rating of 1500. Then we must calculate the expected probability of team A winning the game $p_A$:

\begin{equation*}
    p_A = \frac{1}{1+10^{(e_A - R_B) / 400}}
\end{equation*}

where $R_A$ and $R_B$ are the pre-match Elo ratings of team A an team B, respectively. We note here that $p_B$ is equal to $1-p_A$ as there are no ties in college football.

Once the match has occurred, the rating for team A is updated $(R_A^*)$ by the following:

\begin{equation*}
    R_A^* = R_A + 25 (O_A - p_A)
\end{equation*}

where $O_A$ is the outcome of the match from the perspective of team A, i.e. 1 if they won the match or 0 if they lost. The same calculation is used to adjust the rating of team B:

\begin{equation*}
    R_B^* = R_B + 25 (O_B - p_B)
\end{equation*}

This means that the rating of a team with a high expected probability of winning a match, will not improve by a significant amount, however if a team causes an upset then their rating will increase noticeably. The chosen constant of 25 can be changed in order to make the ratings more or less sensitive to the outcome of single matches.

\section{Results and Discussion}

\subsection{The 2023 College Football Season}

Table~\ref{ratings_2023} shows the Elo ratings of the highest ranked College Football teams on selection day, prior to the CFP committee's selection. According to Elo ratings, the selection committee was justified in excluding Florida State from the 2023 CFP, as they were only the 11th best team in the country at that time. The ratings did not agree with the selection committee any further however as the four teams selected ranked first (Michigan), fifth (Texas), sixth (Alabama), and thirteenth (Washington), making it the season with the largest disagreement between the CFP selection committee and the Elo ratings. The ratings suggest that joining Michigan in the CFP should have been Georgia (the defending national champion), Ohio State and Penn State, both from the Big Ten conference. The ratings for 2023 represent a scenario that the selection committee are highly unlikely to permit based on their criteria--a team with two losses for the season being included alongside the two teams that they have lost to. However, as of selection day, Penn State, while being only the third best team in the Big Ten, were also the forth best team in the country. (Interestingly, this marks the second time that Penn State were ranked in the top 4 according to Elo ratings, but missed selection for the CFP--See Table~\ref{ratings_2017}).

Similarly unlucky in 2023 were Georgia, who despite losing the SEC conference championship to Alabama, were the second best team on selection day according to Elo ratings and were not selected to defend their 2022 national championship.

\begin{table}[htp!]
\centering \caption{The Elo Ratings for the top 13 teams in 2023 College Football prior to the CFP selection.}
\begin{tabular}{c|c|c|c|c}
Elo ranking & Team & Conference & Elo rating & CFP ranking \\
\hline
\rowcolor{Gray}
1 & Michigan & Big Ten & 2174 & 1 \\
2 & Georgia & SEC & 2111 &  \\
3 & Ohio State & Big Ten & 2108 &  \\
4 & Penn State & Big Ten & 2061 &  \\
\rowcolor{Gray}
5 & Texas & Big 12 & 2050 & 3 \\
\rowcolor{Gray}
6 & Alabama & SEC & 2039 & 4 \\
7 & Oregon & Pac-12 & 2031 &  \\
8 & Notre Dame & Independent & 2009 &  \\
9 & LSU & SEC & 1995 &  \\
10 & Oklahoma & Big 12 & 1974 &  \\
11 & Florida State & ACC & 1951 &  \\
12 & Kansas State & Big 12 & 1942 &  \\
\rowcolor{Gray}
13 & Washington & Pac-12 & 1883 & 2 \\
\end{tabular}
\label{ratings_2023}
\end{table}

\subsection{The 2014-2022 College Football Seasons}

Tables~\ref{ratings_2022}--\ref{ratings_2014} show the Elo ratings and the teams selected for the CFP by the committee for each season since the playoff began.

% never chosen the same top 4, twice chosen 4 of the top 5 (2021, 2016)
An inspection of these Tables shows that the CFP committee has never selected the four highest ranked teams according to Elo. However, twice they have selected four of the top five--in 2021 and 2016. The only time the number one ranked team was not selected was Alabama in 2022. There have been three occasions where a selected team was not ranked in the Elo top ten--Notre Dame in 2020 (Elo ranking of 11th), TCU in 2022 (12th), and Washington in 2023 (13th).

\begin{table}[htp!]
\centering \caption{The Elo Ratings for the top 12 teams in 2022 College Football prior to the CFP selection.}
\begin{tabular}{c|c|c|c|c}
Elo ranking & Team & Conference & Elo rating & CFP ranking \\
\hline
1 & Alabama & SEC & 2151 &  \\
\rowcolor{Gray}
2 & Michigan & Big Ten & 2150 & 2 \\
\rowcolor{Gray}
3 & Georgia & SEC & 2146 & 1 \\
\rowcolor{Gray}
4 & Ohio State & Big Ten & 2108 & 4 \\
5 & Penn State & Big Ten & 1979 &  \\
6 & Tennessee & SEC & 1977 &  \\
7 & Utah & Pac-12 & 1936 &  \\
8 & Texas & Big 12 & 1931 &  \\
9 & Kansas State & Big 12 & 1891 &  \\
10 & Clemson & ACC & 1866 &  \\
11 & Notre Dame & Independent & 1858 &  \\
\rowcolor{Gray}
12 & TCU & Big 12 & 1856 & 3 \\
\end{tabular}
\label{ratings_2022}
\end{table}

\begin{table}[htp!]
\centering \caption{The Elo Ratings for the top 10 teams in 2021 College Football prior to the CFP selection.}
\begin{tabular}{c|c|c|c|c}
Elo ranking & Team & Conference & Elo rating & CFP ranking \\
\hline
\rowcolor{Gray}
1 & Alabama & SEC & 2220 & 1 \\
2 & Ohio State & Big Ten & 2153 &  \\
\rowcolor{Gray}
3 & Georgia & SEC & 2143 & 3 \\
\rowcolor{Gray}
4 & Michigan & Big Ten & 2065 & 2 \\
\rowcolor{Gray}
5 & Cincinatti & AAC & 2059 & 4 \\
6 & Notre Dame & Independent & 1992 &  \\
7 & Iowa State & Big 12 & 1929 &  \\
8 & Oklahoma State & Big 12 & 1911 &  \\
9 & Clemson & ACC & 1889 &  \\
10 & Wisconsin & Big Ten & 1870 &  \\
\end{tabular}
\label{ratings_2021}
\end{table}

\begin{table}[htp!]
\centering \caption{The Elo Ratings for the top 11 teams in 2020 College Football prior to the CFP selection.}
\begin{tabular}{c|c|c|c|c}
Elo ranking & Team & Conference & Elo rating & CFP ranking \\
\hline
\rowcolor{Gray}
1 & Clemson & ACC & 2396 & 2 \\
\rowcolor{Gray}
2 & Ohio State & Big Ten & 2373 & 3 \\
3 & LSU & SEC & 2207 &  \\
\rowcolor{Gray}
4 & Alabama & SEC & 2160 & 1 \\
5 & Wisconsin & Big Ten & 2020 &  \\
6 & Oregon & Pac-12 & 2011 &  \\
7 & Georgia & SEC & 1992 &  \\
8 & Michigan & Big Ten & 1984 &  \\
9 & Oklahoma & Big 12 & 1984 &  \\
10 & Penn State & Big Ten & 1968 &  \\
\rowcolor{Gray}
11 & Notre Dame & Independent & 1965 & 4 \\
\end{tabular}
\label{ratings_2020}
\end{table}

\begin{table}[htp!]
\centering \caption{The Elo Ratings for the top 10 teams in 2019 College Football prior to the CFP selection.}
\begin{tabular}{c|c|c|c|c}
Elo ranking & Team & Conference & Elo rating & CFP ranking \\
\hline
\rowcolor{Gray}
1 & Clemson & ACC & 2396 & 3 \\
\rowcolor{Gray}
2 & Ohio State & Big Ten & 2373 & 2 \\
\rowcolor{Gray}
3 & LSU & SEC & 2207 & 1 \\
4 & Alabama & SEC & 2160 &  \\
5 & Wisconsin & Big Ten & 2020 &  \\
6 & Oregon & Pac-12 & 2011 &  \\
7 & Georgia & SEC & 1992 &  \\
8 & Michigan & Big Ten & 1984 &  \\
\rowcolor{Gray}
9 & Oklahoma & Big 12 & 1984 & 4 \\
10 & Penn State & Big Ten & 1968 &  \\
\end{tabular}
\label{ratings_2019}
\end{table}

\begin{table}[htp!]
\centering \caption{The Elo Ratings for the top 10 teams in 2018 College Football prior to the CFP selection.}
\begin{tabular}{c|c|c|c|c}
Elo ranking & Team & Conference & Elo rating & CFP ranking \\
\hline
\rowcolor{Gray}
1 & Alabama & SEC & 2324 & 1 \\
\rowcolor{Gray}
2 & Clemson & ACC & 2277 & 2 \\
3 & Georgia & SEC & 2035 &  \\
\rowcolor{Gray}
4 & Oklahoma & Big 12 & 2026 & 4 \\
5 & Ohio State & Big Ten & 2018 &  \\
6 & Michigan & Big Ten & 1966 &  \\
\rowcolor{Gray}
7 & Notre Dame & Independent & 1953 & 3 \\
8 & Mississippi State & SEC & 1949 &  \\
9 & Penn State & Big Ten & 1929 &  \\
10 & Iowa & Big Ten & 1926 &  \\
\end{tabular}
\label{ratings_2018}
\end{table}

\begin{table}[htp!]
\centering \caption{The Elo Ratings for the top 10 teams in 2017 College Football prior to the CFP selection.}
\begin{tabular}{c|c|c|c|c}
Elo ranking & Team & Conference & Elo rating & CFP ranking \\
\hline
\rowcolor{Gray}
1 & Alabama & SEC & 2206 & 1 \\
\rowcolor{Gray}
2 & Clemson & ACC & 2141 & 2 \\
3 & Penn State & Big Ten & 2126 &  \\
\rowcolor{Gray}
4 & Ohio State & Big Ten & 2104 & 3 \\
5 & Oklahoma & Big 12 & 2070 &  \\
6 & Georgia & SEC & 2040 &  \\
7 & Wisconsin & Big Ten & 2026 &  \\
\rowcolor{Gray}
8 & Washington & Pac-12 & 2017 & 4 \\
9 & Auburn & SEC & 2015 &  \\
10 & Oklahoma State & Big 12 & 1899 &  \\
\end{tabular}
\label{ratings_2017}
\end{table}

\begin{table}[htp!]
\centering \caption{The Elo Ratings for the top 10 teams in 2016 College Football prior to the CFP selection.}
\begin{tabular}{c|c|c|c|c}
Elo ranking & Team & Conference & Elo rating & CFP ranking \\
\hline
\rowcolor{Gray}
1 & Ohio State & Big Ten & 2328 & 3 \\
\rowcolor{Gray}
2 & Alabama & SEC & 2311 & 1 \\
\rowcolor{Gray}
3 & Washington & Pac-12 & 2149 & 4 \\
4 & Michigan & Big Ten & 2142 &  \\
\rowcolor{Gray}
5 & Clemson & ACC & 2077 & 2 \\
6 & Oklahoma & Big 12 & 2013 &  \\
7 & USC & Pac-12 & 1951 &  \\
8 & Western Kentucky & Conference USA & 1941 &  \\
9 & Auburn & SEC & 1938 &  \\
10 & LSU & SEC & 1932 &  \\
\end{tabular}
\label{ratings_2016}
\end{table}

\begin{table}[htp!]
\centering \caption{The Elo Ratings for the top 10 teams in 2015 College Football prior to the CFP selection.}
\begin{tabular}{c|c|c|c|c}
Elo ranking & Team & Conference & Elo rating & CFP ranking \\
\hline
\rowcolor{Gray}
1 & Oklahoma & Big 12 & 2155 & 4 \\
2 & Ohio State & Big Ten & 2107 &  \\
\rowcolor{Gray}
3 & Alabama & SEC & 2106 & 2 \\
4 & Baylor & Big 12 & 1975 &  \\
5 & Stanford & Pac-12 & 1967 &  \\
\rowcolor{Gray}
6 & Clemson & ACC & 1965 & 1 \\
\rowcolor{Gray}
7 & Michigan State & Big Ten & 1944 & 3 \\
8 & TCU & Big 12 & 1943 &  \\
9 & Florida State & ACC & 1938 &  \\
10 & Houston & AAC & 1891 &  \\
\end{tabular}
\label{ratings_2015}
\end{table}

\begin{table}[htp!]
\centering \caption{The Elo Ratings for the top 10 teams in 2014 College Football prior to the CFP selection.}
\begin{tabular}{c|c|c|c|c}
Elo ranking & Team & Conference & Elo rating & CFP ranking \\
\hline
\rowcolor{Gray}
1 & Ohio State & Big Ten & 2232 & 4 \\
\rowcolor{Gray}
2 & Alabama & SEC & 2139 & 1 \\
\rowcolor{Gray}
3 & Oregon & Pac-12 & 2119 & 2 \\
4 & TCU & Big 12 & 2105 &  \\
5 & Baylor & Big 12 & 2061 &  \\
6 & Michigan State & Big Ten & 2025 &  \\
\rowcolor{Gray}
7 & Florida State & ACC & 2012 & 3 \\
8 & Georgia & SEC & 2004 &  \\
9 & Mississippi State & SEC & 1933 &  \\
10 & Kansas State & Big 12 & 1891 &  \\
\end{tabular}
\label{ratings_2014}
\end{table}

\subsection{Should Elo replace the CFP selection committee?}

Whether the CFP selection committee should be replaced by an Elo ratings method is difficult to answer for anyone outside of the CFP organization.
If the goal of the CFP is to have transparency befitting of the elite-level on-field product, then an Elo model is a far better choice.
Elo takes into account strength of schedule and compares outcomes to expected outcomes, which are all known quantities and can be calculated by teams hoping to qualify.
Elo also has a heavier weighting on recent matches, ensuring that the teams qualifying for the CFP are currently the best four choices.
Overall, an Elo model removes subjectivity, but this might not be the purpose of the CFP selection.
If the intention of the CFP is to draw the most viewers and media attention then the subjectivity offered by employing the CFP selection committee is the best way to achieve this.

\subsection{Future CFPs}

Starting in the 2024 season, the CFP will be expanded to include 12 teams competing for the national championship.
However, the CFP expansion will retain the current selection-based criteria, meaning that rather than discussing the `unlucky' fifth team who missed out, the sports media will discuss the `unlucky' thirteenth team.
As above, if the CFP process wants to be transparent, a ratings model would be the only way to achieve this.

\section{Conclusion}
In this paper we compared the teams chosen by the CFP committee to compete for the NCAA college football national championship with an Elo ratings method. We found that the top four teams by Elo ratings differ greatly from the four teams selected in almost every year of the CFP's existence. We use this to suggest that the subjective criteria used by the selection committee should be replaced by a ratings model in the name of transparency.


\begin{thebibliography}{18}
    \providecommand{\natexlab}[1]{#1}
    \providecommand{\url}[1]{\texttt{#1}}
    \expandafter\ifx\csname urlstyle\endcsname\relax
      \providecommand{\doi}[1]{doi: #1}\else
      \providecommand{\doi}{doi: \begingroup \urlstyle{rm}\Url}\fi
    
    \bibitem[Angelini et~al.(2022)Angelini, Candila, and De~Angelis]{angelini2022weighted}
    Giovanni Angelini, Vincenzo Candila, and Luca De~Angelis.
    \newblock Weighted {E}lo rating for tennis match predictions.
    \newblock \emph{European Journal of Operational Research}, 297\penalty0 (1):\penalty0 120--132, 2022.
    
    \bibitem[Arntzen and Hvattum(2021)]{arntzen2021predicting}
    Halvard Arntzen and Lars~Magnus Hvattum.
    \newblock Predicting match outcomes in association football using team ratings and player ratings.
    \newblock \emph{Statistical Modelling}, 21\penalty0 (5):\penalty0 449--470, 2021.
    
    \bibitem[Collingwood et~al.(2022)Collingwood, Wright, and Brooks]{collingwood2022evaluating}
    James~AP Collingwood, Michael Wright, and Roger~J Brooks.
    \newblock Evaluating the effectiveness of different player rating systems in predicting the results of professional snooker matches.
    \newblock \emph{European Journal of Operational Research}, 296\penalty0 (3):\penalty0 1025--1035, 2022.
    
    \bibitem[Dabadghao and Vaziri(2022)]{dabadghao2022predictive}
    SS~Dabadghao and B~Vaziri.
    \newblock The predictive power of popular sports ranking methods in the {NFL}, {NBA}, and {NHL}.
    \newblock \emph{Operational Research}, 22\penalty0 (3):\penalty0 2767--2783, 2022.
    
    \bibitem[ESPN(2023)]{espnFloridaState}
    ESPN.
    \newblock '{D}isgusted, infuriated': 13-0 {F}{S}{U} snubbed by {C}{F}{P}.
    \newblock \url{https://www.espn.com/college-football/story/_/id/39034100/undefeated-florida-state-left-college-football-playoff}, 2023.
    
    \bibitem[G{\'a}squez and Royuela(2016)]{gasquez2016determinants}
    Roberto G{\'a}squez and Vicente Royuela.
    \newblock The determinants of international football success: A panel data analysis of the {E}lo rating.
    \newblock \emph{Social Science Quarterly}, 97\penalty0 (2):\penalty0 125--141, 2016.
    
    \bibitem[Hvattum and Arntzen(2010)]{hvattum2010using}
    Lars~Magnus Hvattum and Halvard Arntzen.
    \newblock Using {E}lo ratings for match result prediction in association football.
    \newblock \emph{International Journal of forecasting}, 26\penalty0 (3):\penalty0 460--470, 2010.
    
    \bibitem[Lee et~al.(2018)Lee, Danileiko, and Vi]{lee2018testing}
    Michael~D Lee, Irina Danileiko, and Julie Vi.
    \newblock Testing the ability of the surprisingly popular method to predict {NFL} games.
    \newblock \emph{Judgment and Decision Making}, 13\penalty0 (4):\penalty0 322--333, 2018.
    
    \bibitem[Morse(2023)]{cnnFloridaState}
    Ben Morse.
    \newblock {W}hy {F}lorida {S}tate was left out of the {C}ollege {F}ootball {P}layoff and why it’s so controversial.
    \newblock \url{https://www.cnn.com/2023/12/04/sport/florida-state-left-out-college-football-playoffs-explainer-spt-intl/index.html}, 2023.
    
    \bibitem[Pel{\'a}nek(2016)]{pelanek2016applications}
    Radek Pel{\'a}nek.
    \newblock Applications of the {E}lo rating system in adaptive educational systems.
    \newblock \emph{Computers \& Education}, 98:\penalty0 169--179, 2016.
    
    \bibitem[Playoff(2016)]{criteria}
    College~Football Playoff.
    \newblock {C}{F}{P} {S}election {C}ommittee {P}rotocol.
    \newblock \url{https://collegefootballplayoff.com/sports/2016/10/24/selection-committee-protocol}, 2016.
    
    \bibitem[Playoff(2017)]{committee}
    College~Football Playoff.
    \newblock {C}{F}{P} {S}election {C}ommittee.
    \newblock \url{https://collegefootballplayoff.com/sports/2017/10/16/selection-committee.aspx}, 2017.
    
    \bibitem[Ryall and Bedford(2010)]{ryall2010optimized}
    Richard Ryall and Anthony Bedford.
    \newblock An optimized ratings-based model for forecasting {A}ustralian {R}ules football.
    \newblock \emph{International Journal of Forecasting}, 26\penalty0 (3):\penalty0 511--517, 2010.
    
    \bibitem[sports(2023)]{cbssportsFloridaState}
    CBS sports.
    \newblock {F}lorida {S}tate {A}{D} blasts {C}ollege {F}ootball {P}layoff after being snubbed from four-team field in favor of {A}labama.
    \newblock \url{https://www.cbssports.com/college-football/news/florida-state-ad-blasts-college-football-playoff-after-being-snubbed-from-four-team-field-in-favor-of-alabama/}, 2023.
    
    \bibitem[Tenkanen(2019)]{tenkanen2019rating}
    Santeri Tenkanen.
    \newblock Rating {N}ational {H}ockey {L}eague teams: the predictive power of {E}lo rating models.
    \newblock \emph{International Journal of Forecasting}, 22\penalty0 (4):\penalty0 679--688, 2019.
    
    \bibitem[Tsang et~al.(2016)Tsang, Ngan, and Pang]{tsang2016fabric}
    Colin~SC Tsang, Henry~YT Ngan, and Grantham~KH Pang.
    \newblock Fabric inspection based on the {E}lo rating method.
    \newblock \emph{Pattern Recognition}, 51:\penalty0 378--394, 2016.
    
    \bibitem[Vaughan~Williams et~al.(2021)Vaughan~Williams, Liu, Dixon, and Gerrard]{vaughan2021well}
    Leighton Vaughan~Williams, Chunping Liu, Lerato Dixon, and Hannah Gerrard.
    \newblock How well do {E}lo-based ratings predict professional tennis matches?
    \newblock \emph{Journal of Quantitative Analysis in Sports}, 17\penalty0 (2):\penalty0 91--105, 2021.
    
    \bibitem[Ziemba(2015)]{ziemba20152014}
    Bill Ziemba.
    \newblock The 2014--2015 {{NFL}} season, playoffs, and the super bowl.
    \newblock \emph{Wilmott}, 2015\penalty0 (77):\penalty0 24--43, 2015.
    
\end{thebibliography}
\end{document}